\documentclass{raa}            % referee version: for submission

\usepackage{graphicx,times}             %for PS/EPS graphics inclusion, new
\usepackage{natbib}
\usepackage{amssymb,amsmath}
\bibpunct{(}{)}{;}{a}{}{,}

\usepackage[a4paper=true,dvipdfm=true,pagebackref=true]{hyperref}
\hypersetup{colorlinks = true, linkcolor = green, anchorcolor = red, citecolor = blue, filecolor = red, pagecolor = red, urlcolor = red}

\begin{document}

   \title{Methanol formation chemistry with revised reactions scheme
%\footnotetext{$*$ Supported by the National Natural Science Foundation of China.}
}
%   \subtitle{I. Place Your Subtitle Here}

   \volnopage{Vol.0 (20xx) No.0, 000--000}      %%preserved for Editor. DOn't remove!
   \setcounter{page}{1}          %%starting page, preserved for Editor. DOn't remove!

   \author{V. A. Sokolova
      \inst{1,2}
      }

   \institute{Ural Federal University, 19 Mira street, Yekaterinburg, 620002, Russia; {\it valeria.sokolova@urfu.ru}\\
        \and
             Engineering Research Institute "Ventspils International Radio Astronomy Centre" of Ventspils University College, Inzenieru 101, Ventspils, LV-3601, Latvia\\
\vs\no
   {\small Received~~20xx month day; accepted~~20xx~~month day}}

\abstract{The aim of the presented work is to analyze the impact of experimentally evaluated reactions of hydrogen abstraction on surfaces of interstellar grains on the chemical evolution of methanol and its precursors on grains and in the gas phase under conditions of cold dark cloud and during the collapse of the translucent cloud into the dark cloud. Analysis of simulation results shows that those reactions are highly efficient destruction channels for HCO and H$_2$CO on grain surfaces, and significantly impact the abundances of almost all molecules participating in the formation of CH$_3$OH. Next, in models with those reactions maximum abundances of methanol in gas and on grain surface decrease by more than 2-3 orders of magnitude in comparison to models without surface abstraction reactions of hydrogen. Finally, we study the impact of binding energies of CH$_2$OH and CH$_3$O radicals on methanol chemistry.
}

   \authorrunning{V. A. Sokolova}            %author_head in even pages
   \titlerunning{Methanol formation chemistry with revised reactions scheme }  % title_head in odd pages

   \maketitle
%% The author head (on even pages) and the title head (on odd pages) will be
%% automatically extracted from \author{} and \title{}. Whenever the title is too long,
%% you will be asked to supply a shorter one by inserting either \authorrunning{} or
%% \titlerunning{} before \maketitle. Anyway, you can specify your own heads.
%%
%%
%% Note: In the following text body of your manuscript, please note several differences from
%%       other major journals:
%% (1) \subsection{Please Capitalize the First Letter of Each Notional Word in Subsection Title}
%% (2) Please Capitalize the First Letter of Each Notional Word in all tables' captions

%
%________________________________________________ sections below
%
\section{Introduction}           %% first-level sections will be auto-capitalized
\label{sect:intro}
Diverse molecular composition is one of the main properties of interstellar objects: it can include both simple and complex (more than 6 atoms) molecules (\cite{2018ApJS..239...17M}). Complex organic molecules (COMs, molecules containing more than 6 atoms, including C and H) are species of special interest, because they are actively involved in prebiotic chemistry and are associated with the origin of life (\cite{2009ARA&A..47..427H}). Methanol (CH$_{3}$OH) is important molecule for the formation of more complex organic compounds according to a number of laboratory and theoretical studies (\cite{2009A&A...504..891O, murga, rivilla, kochina}). It is assumed now (see e. g.~\cite{2002ApJ...571L.173W, 2009A&A...505..629F, 2015arXiv150702729L}) that methanol in ISM forms on the surface of dust particles by hydrogenation of a carbon monoxide (CO) molecule, and then desorbs into the gas phase via thermal and non-thermal processes. Conventional hydrogenation that occurs on grain surfaces and leads from CO to CH$_3$OH (\cite{1991ApJ...381..181T}) is as follows:
\begin{equation}
\text{CO}  \xrightarrow[  ]{+ H} \text{HCO} \xrightarrow[  ]{+ H} \text{H}_{2}\text{CO} \xrightarrow[  ]{+ H} \text{CH}_{2}\text{OH} \xrightarrow[  ]{+ H} \text{CH}_{3}\text{OH}.\label{COH}
\end{equation}

Recent laboratory experiments by~\cite{Mini2016} on atomic hydrogen exposure of carbon monoxide, formaldehyde, and methanol thin films on cold surfaces revealed an unexpected desorption phenomenon. The analysis of experiments led the authors to the conclusion that hydrogenation sequence of CO on grain surface must be expanded. It was shown that it is necessary to include additional H$_{2}$-abstraction reactions (reverse reactions) for HCO and H$_{2}$CO into the scheme of surface methanol formation:
\begin{gather}
\text{HCO + H} \rightarrow \text{CO + H}_{2}, \label{eq: reaction1}\\
\text{H}_{2}\text{CO + H} \rightarrow \text{HCO + H}_{2}\label{eq: reaction2}.
\end{gather}

Given the fundamental importance of methanol for interstellar chemistry of complex organic molecules, we decided to study the impact of these reactions on chemical evolution under different conditions of interstellar medium. This paper is organized as follows. Section 2 is dedicated to utilized method of computation of molecular abundances and utilized physical models. Section 3 describes the modelling results. Discussion of obtained modelling results is presented in Section 4.

\section{Methods} \label{sect:methods}
\subsection{Calculation molecular abundances}

We use the MONACO code described in~\cite{Vas2017} to calculate abundances of species. It utilizes the rate equations method to numerically simulate the chemical evolution of the interstellar medium, and includes treatment of chemistry in the gas phase and on grain surfaces under non-stationary conditions. This code calculates time-dependent fractional abundances of species with respect to the total number of hydrogen nuclei for each species on a given time interval. In our study we run two types of models: with and without reverse reactions for HCO and H$_2$CO from~\cite{Mini2016}, where the reaction~\eqref{eq: reaction2} has the activation barrier E = 202~K~(\cite{Baulch2005}).

In all models we use kinetic database utilized in~\cite{Vas2017}, with binding energies  E$_d$(CH$_2$OH) = 5\,080~K and  E$_d$(CH$_3$O) = 2\,540~K (see Table 2 in~\cite{Wakelam2017}). We denote the model without reverse reactions as Model I, and model with reverse reactions as Model II.

\subsection{Utilized physical models}

In our analysis we utilized two physical models. First model is a 0D model with time-dependent physical conditions that mimics the transition from a translucent cloud to a dark cloud. Second model represents typical static dark molecular cloud.

The model of collapse was taken from~\cite{VasHerbst2013_2} and consists of two stages. At the first "cold" stage the collapse proceeds in a free-fall regime and starts at a gas density n$_H$ = $3 \times 10^3$ cm$^{-3}$ and visual extinction A$_V$ = 2$^m$. The process continues until a density of $10^7$ cm$^{-3}$ is reached within 10$^6$ yrs of evolution. During the cold stage temperature linearly decreases from 20 to 10~K, and the gas temperature is assumed to be equal to the dust temperature. At the second "warm-up" stage temperature is growing from 10 to 200~K with a square-law over
$2\times10^{5}$ years, that corresponds to models of the formation of intermediate-mass stars. Gas and dust temperatures are equal. Density and visual extinction remain constant. "High-metal" atomic initial composition corresponds to the values listed in column EA2 in Table 1 from~\cite{Wakelam2008}.

Model of a cold dark cloud corresponds to the model described in~\cite{VasHerbst2013_1} and represents a typical cold dark molecular cloud: T = 10~K, proton density n$_H$ = $10^5$ cm$^{-3}$, visual extinction A$_V$ = 10$^m$ and "low-metal" initial composition (see~\cite{Wakelam2008}, column EA1 in Table 1). Chemical evolution is simulated over a time span of $10^6$ years.

\section{Results}

\begin{figure*}
    \includegraphics[width=0.9\textwidth]{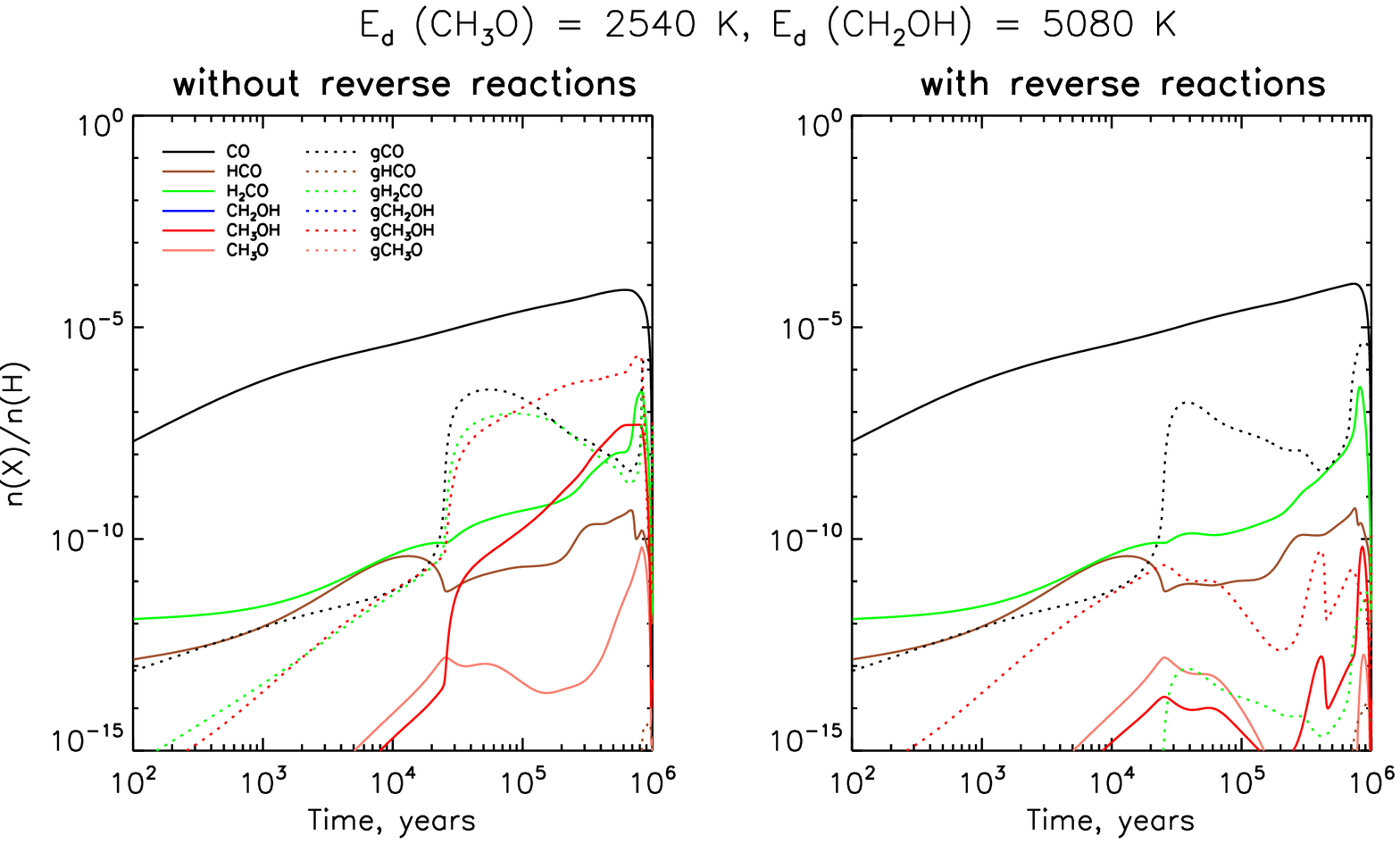}
    \caption{Relative abundances of species versus time for the "cold" stage of the collapse model. These plots correspond to calculations with E$_d$(CH$_2$OH) = 5\,080~K and  E$_d$(CH$_3$O) = 2\,540~K. Solid line represents molecules in gas phase, dashed line~--- molecules on grain surface. The evolution time range is 10$^2$---10$^6$ years.}
    \label{fig:collapse_cold}
\end{figure*}

\begin{figure*}
    \includegraphics[width=0.9\textwidth]{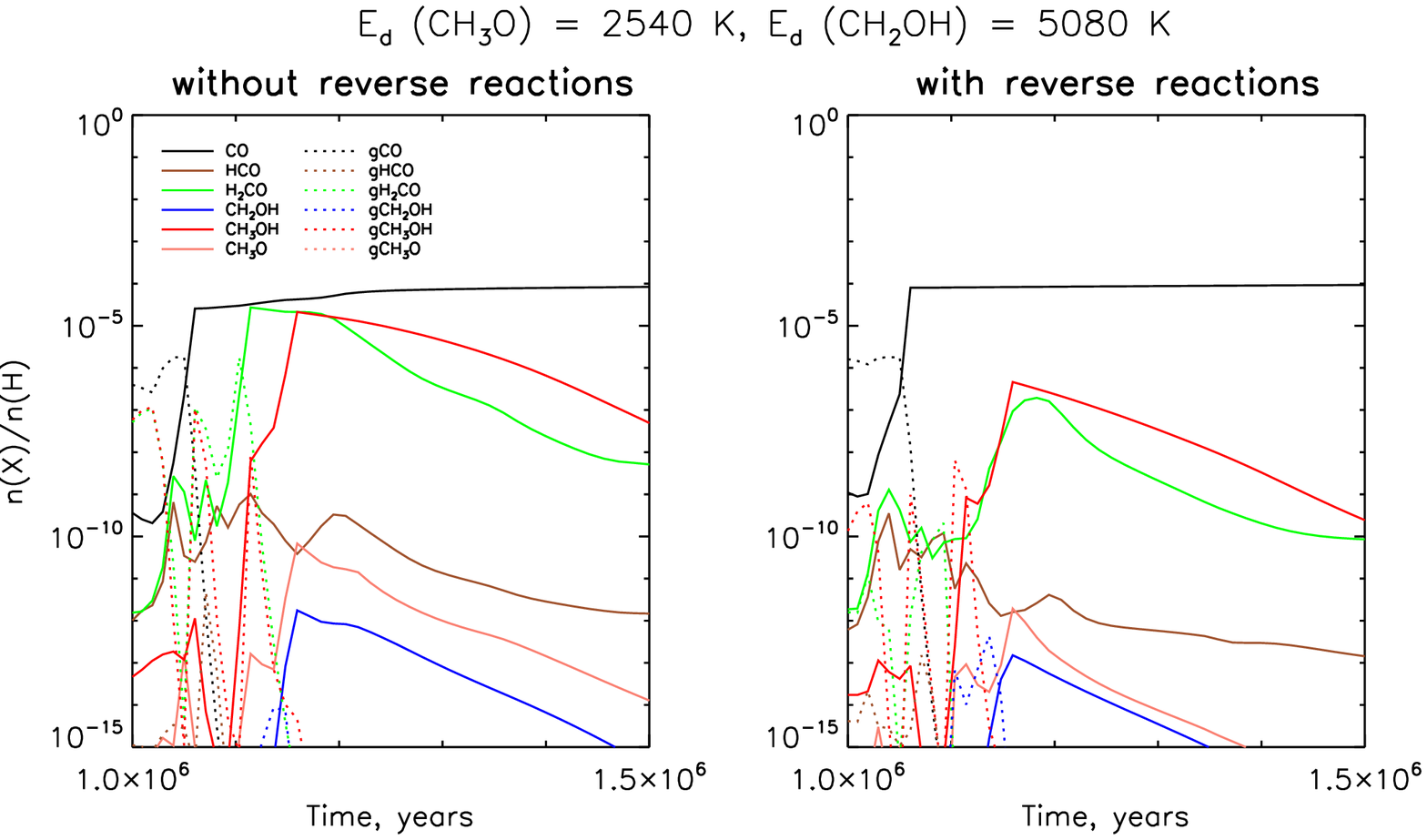}
    \caption{Relative abundances of species versus time for the "warm-up" stage of the collapse model. These plots correspond to calculations with E$_d$(CH$_2$OH) = 5\,080~K and  E$_d$(CH$_3$O) = 2\,540~K. Solid line represents molecules in gas phase, dashed line~--- molecules on grain surface. The evolution time range is 1---1.5$\times$10$^6$ years.}
    \label{fig:collapse_warm}
\end{figure*}

The results of simulations are shown in Figures~\ref{fig:coldcloud}~---~\ref{fig:collapse_warm}. Each plot represents time dependent abundances of methanol and some chemically related species.

\subsection{Cold Cloud}\label{subsec: coldcloud}

\begin{figure*}
    \includegraphics[width=0.9\textwidth]{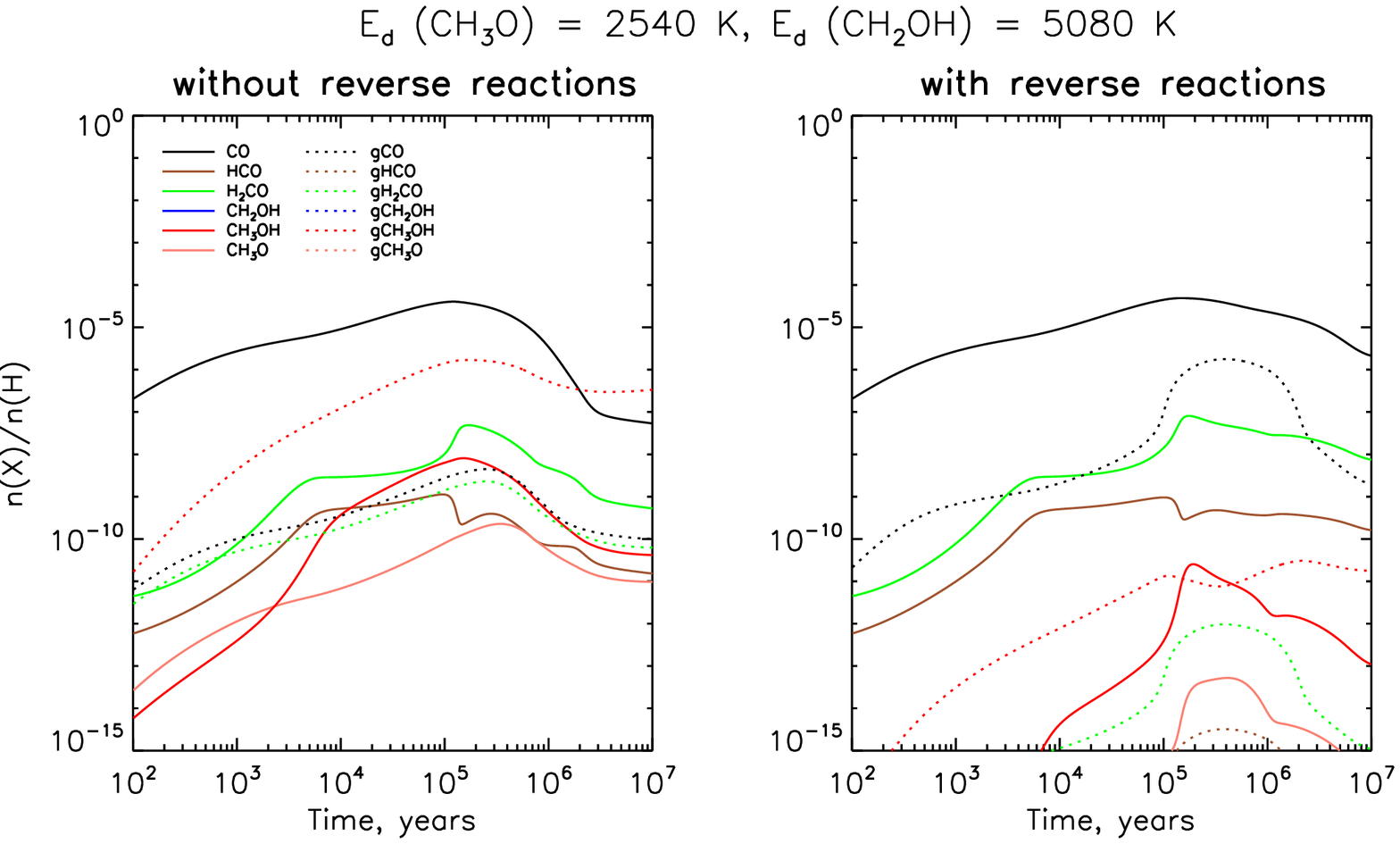}
    \caption{Relative abundances of species versus time for the model of cold dark cloud. These plots correspond to calculations with E$_d$(CH$_2$OH) = 5\,080~K and  E$_d$(CH$_3$O) = 2\,540~K. Solid line represents molecules in gas phase, dashed line~--- molecules on grain surface. The evolution time range is 10$^2$---10$^6$ years.}
    \label{fig:coldcloud}
\end{figure*}

In Figure~\ref{fig:coldcloud} the abundances of of methanol and chemically related species as a function of time in the gas phase and on the grain surface for the model of the cold dark cloud are presented.

In the Model II abundances of CH$_{3}$OH (both gaseous and on grain surface), gaseous CH$_2$OH, H$_2$CO, HCO and CO on grain surface show significant changes in comparison with the Model I.

In the Model I, the main methanol formation routes are through the following reactions on the grain surface followed by the efficient reactive desorption (here and after in formulas and in figures all species on grain surface are marked with prefix 'g', species without 'g'-prefix are the gaseous species)
\begin{gather}
\text{gH + gCH}_{3}\text{O} \rightarrow \text{CH}_{3}\text{OH}, \label{eq: reaction3}\\
\text{gH + gCH}_{2}\text{OH} \rightarrow \text{CH}_{3}\text{OH}, \label{eq: reaction4}
\end{gather}
where the rate of the reaction~\eqref{eq: reaction3} is higher by the factor of three than in reaction~\eqref{eq: reaction4}. In the Model II, reactions~\eqref{eq: reaction3} and~\eqref{eq: reaction4} become inefficient because of strong backward reactions in the nets of CH$_3$O and CH$_2$OH formation. Dissociative recombination of gaseous H$_2$COHOCH$_2^+$ becomes the main route of gas phase methanol formation
\begin{equation}\label{eq: reaction5}
\text{H}_{2}\text{COHOCH}_{2}^{+} + e^- \rightarrow \text{CH}_{3}\text{OH + HCO}.
\end{equation}

Molecular ion H$_2$COHOCH$_2^+$ in the Model II forms in reactions
\begin{gather}
\text{CH}_{3}^{+} + \text{CH}_{2}\text{O}_{2}\rightarrow \text{H}_{2}\text{COHOCH}_{2}^{+}, \label{eq:ext2}\\
\text{H}_{3}\text{CO}^{+} + \text{H}_{2}\text{CO} \rightarrow \text{H}_{2}\text{COHOCH}_{2}^{+}. \label{eq:ext3}
\end{gather}

Analysis of chemical pathways shows that in both models the major reactions for CH$_2$OH and CH$_3$O formation on grain surface are
\begin{gather}
\text{gH + gH}_{2}\text{CO} \rightarrow \text{gCH}_{2}\text{OH}, \label{eq: reaction6}\\
\text{gH + gH}_{2}\text{CO} \rightarrow \text{gCH}_{3}\text{O}\label{eq: reaction7}
\end{gather}
with equal rates. It means that smaller binding energy of CH$_3$O affects its abundance (it becomes higher) and helps CH$_3$O molecule to replace CH$_2$OH in reactions described above. At the same time, in the Model II the rate of reaction~\eqref{eq: reaction2} is high enough to quickly destroy H$_2$CO. This is the reason of high HCO abundance on grain surface too. One should note that behaviour of HCO and CO maximum abundances on grain surface (see Figure~\eqref{fig:coldcloud}) are similar to each other and this can be also explained by high rate of reaction~\eqref{eq: reaction2}.

Methanol on grain surfaces in both models is mainly formed by reactions
\begin{gather}
\text{gH + gCH}_{2}\text{OH} \rightarrow \text{gCH}_{3}\text{OH},\label{eq: reaction8}\\
\text{gH + gCH}_{3}\text{O} \rightarrow \text{gCH}_{3}\text{OH}. \label{eq: reaction9}
\end{gather}
In the Model II efficiencies of reactions~\eqref{eq: reaction8} and ~\eqref{eq: reaction9} decreases and accretion of CH$_3$OH from gas becomes important in chemical evolution of methanol on grain surface.

Gaseous CH$_2$OH in the Model I can be formed only by reaction
\begin{equation}
\text{gH + gH}_{2}\text{CO} \rightarrow \text{CH}_{2}\text{OH}. \label{eq: reaction10}
\end{equation}
The rate of this reaction in the Model II decreases by the factor of five and the main route of CH$_2$OH formation in gas is
\begin{equation}
\text{gOH + gCH}_2 \rightarrow \text{CH}_2\text{OH} \label{eq: reaction11}.
\end{equation}

Gaseous CH$_3$O in the Model I is formed in reactions
\begin{gather}
\text{gH + gH}_{2}\text{CO} \rightarrow \text{CH}_{3}\text{O}, \label{eq: reaction12} \\
\text{CH}_3\text{OH + OH} \rightarrow \text{CH}_3\text{O + H}_2\text{O}. \label{eq: reaction13}
\end{gather}

Reaction~\eqref{eq: reaction13} becomes the most effective route of CH$_3$O formation and rate of reaction~\eqref{eq: reaction12} drops down to zero in the Model II.

For CH$_2$OH and CH$_3$O, and then for CH$_3$OH, such changes in abundances and formation pathways can be explained by decreasing H$_2$CO abundance on grain surface as it was mentioned above.

\subsection{Collapse from the translucent cloud into the dark cloud}\label{subsec: collapse}
\subsubsection{Cold stage}\label{subsubsec: col_1}
Abundances of molecules in the gas and on grain surface for the first stage of the model of collapse are presented in Figure~\ref{fig:collapse_cold}. Molecules with significant changes in abundances are the same as in the model of cold dark cloud.

In both types of models (Model I and II) gaseous methanol is formed in reactions~\eqref{eq: reaction3},~\eqref{eq: reaction4} and released in gas phase by thermal desorption from the grain surface. Rate of reaction~\eqref{eq: reaction6} is higher by the factor of two than in the Model I. In the Model I, the main pathways of CH$_3$O and CH$_2$OH formation are reactions~\eqref{eq: reaction6} and~\eqref{eq: reaction7}, but in the Model II almost all CH$_3$O molecules on grain surface are formed in the reaction
\begin{equation} \label{eq: reaction14}
\text{gO + gCH}_{3} \rightarrow \text{gCH}_{3}\text{O},
\end{equation}
with temperatures of the medium of 18-20~K. At the same temperatures CH$_2$OH is formed in the reaction
\begin{equation} \label{eq: reaction_ext}
\text{gOH + gCH}_{2} \rightarrow \text{gCH}_{2}\text{OH},
\end{equation}
Then during the collapse CH$_3$O begin to form in the reaction~\ref{eq: reaction7}, and CH$_2$OH forms in reactions~\ref{eq: reaction6} and~\ref{eq: reaction_ext}.

As it was described for the case of cold dark cloud, high rate of the reverse reaction~\eqref{eq: reaction2} is responsible for decreasing of H$_2$CO abundance on grain surface and it affects on other reactions with this molecule.

\subsubsection{Warm-up stage}\label{subsubsec: col_1}

Abundances of gaseous molecules and molecules on grains as a function of time for the second "warm" stage of the collapse are shown in Figure~\ref{fig:collapse_warm}. Molecules with significant changes in abundances are the same as in the case of cold stage of the collapse. Warm-up stage is characterized by the square-law growing of temperature. Below we discuss molecular formation chains which occur at times when molecules start to approach their maximum values.

In both models (I and II) formation of CH$_{3}$O and CH$_{2}$OH becomes the object of our interest. Formation pathways of CH$_{3}$O (both gaseous and on the grain surface) are different from those described for the models of cold dark cloud and cold stage of the collapse. Gaseous CH$_{3}$O in both models (I and II) are produced effectively in reactions~\eqref{eq: reaction13} and~\eqref{eq: reaction12}, but reaction~\eqref{eq: reaction12} is less effective than reaction~\eqref{eq: reaction13}. In the Model I CH$_{3}$OH forms in reactions~\eqref{eq: reaction3} and~\eqref{eq: reaction4} with equal efficiencies, but in the model II CH$_{3}$OH forms in reaction~\eqref{eq: reaction4} only. CH$_{2}$OH on grain surface in the model I forms in reaction~\eqref{eq: reaction6}, but in the Model II there is no efficient routes for its formation.

The main pathway of CH$_{3}$O formation on grains in both models is reaction~\eqref{eq: reaction7} and the strongest reaction of its destruction is reaction~\eqref{eq: reaction9}. In the Model II reaction~\eqref{eq: reaction7} becomes inefficient from chemical evolution of this molecule so CH$_{3}$O on the grain surfaces can be destructed only in reaction~\eqref{eq: reaction9}, and there is no routes for its formation. All changes in chemical evolution of gCH$_{3}$O can be explained by the influence of the added reverse reactions. In the Model II one can see that H$_{2}$CO on grain surfaces is destroyed in reaction~\eqref{eq: reaction2} only, and there are no routes of effective CH$_2$OH, CH$_3$O formation on grain surfaces in contrast with Model I. This strongly affects on formation of CH$_{3}$OH, because CH$_{2}$OH on the grain surface is a key reactant in the reaction~\eqref{eq: reaction4}, and thus affects on formation of gaseous CH$_{3}$O through reaction~\eqref{eq: reaction13}. Moreover, rate of H$_{2}$CO destruction on grain surfaces in the Model II is higher by the factor of two than its formation and this directly affects on CH$_{3}$O formation in reaction~\eqref{eq: reaction7} on grain surface.

In both models CH$_{2}$OH in gas and on grain surface is formed less effectively than CH$_{3}$O. Main formation pathways of gaseous CH$_{2}$OH are reaction~\eqref{eq: reaction11} and less effective reaction~\eqref{eq: reaction10}. In the Model II reaction~\eqref{eq: reaction10} inefficient in chemical evolution of gaseous CH$_{2}$OH, so only reaction~\eqref{eq: reaction11} takes part in the formation of this molecule. Network analysis of CH$_{2}$OH formation on grain surfaces shows that in the Model I destruction of CH$_{2}$OH on grain surface in reaction~\eqref{eq: reaction8} proceeds more effectively than its formation in reaction~\eqref{eq: reaction6}. In the Model II, reaction~\eqref{eq: reaction6} disappears from chemical evolution, so CH$_{2}$OH on grains can be destructed only in reaction~\eqref{eq: reaction8}. The most probable explanation is that CH$_2$OH chemical evolution is determined by faster destruction of H$_{2}$CO.

\section{Discussion and conclusions}

In this study we analyzed the impact of added reverse reactions (reactions of H$_2$ abstraction) for HCO and H$_{2}$CO on the grain surfaces proposed by~\cite{Mini2016} on methanol chemistry. Analysis shows that the new reactions are highly efficient in the destruction of HCO and H$_2$CO and have a significant impact on abundances of molecules in the networks of methanol formation. In the models with reverse reactions abundances of CH$_{3}$OH in gas phase and on grain surface decreases by 2-3 orders of magnitude.

These results are in a somewhat mixed agreement with observations. In our model with added reverse reactions fractional abundances of methanol are about 10$^{-11}$ for the model of cold dark cloud and 10$^{-10}$ and 10$^{-6}$ for the first and the second stages of collapse correspondingly. Observed relative abundance of CH$_3$OH in cold dark medium is about $2 \times 10^{-9}$ (\cite{1991Icar...91....2I, 2013ChRv..113.8710A, 2016ApJ...830L...6J, 2014ApJ...780...85V}) and in hot cores - $10^{-7}$---$10^{-5}$ (\cite{2007A&A...463..601B, 2009ARA&A..47..427H}). This discrepancy raises questions about the exact values of activation barriers of reverse reactions for HCO and H$_{2}$CO on grain surfaces, which will be solved in future studying.

\begin{figure*}
    \includegraphics[width=0.9\textwidth]{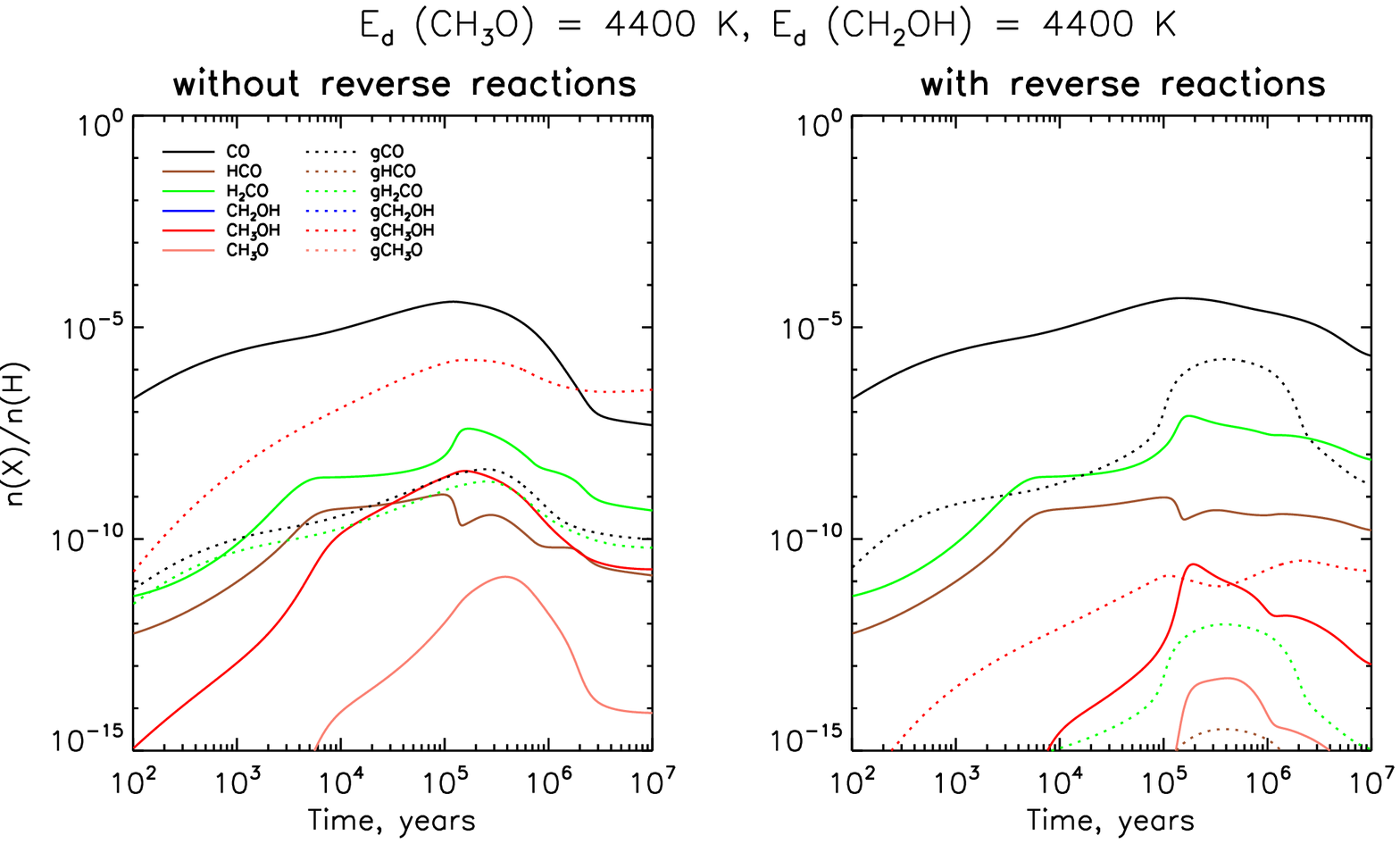}
    \caption{Relative abundances of species versus time for the model of cold dark cloud. These plots correspond to calculations with E$_d$(CH$_2$OH) =  E$_d$(CH$_3$O) = 4\,400~K. Solid line represents molecules in gas phase, dashed line~--- molecules on grain surface. The evolution time range is 10$^2$---10$^6$ years.}
    \label{fig:coldcloud1}
\end{figure*}

\begin{figure*}
    \includegraphics[width=0.9\textwidth]{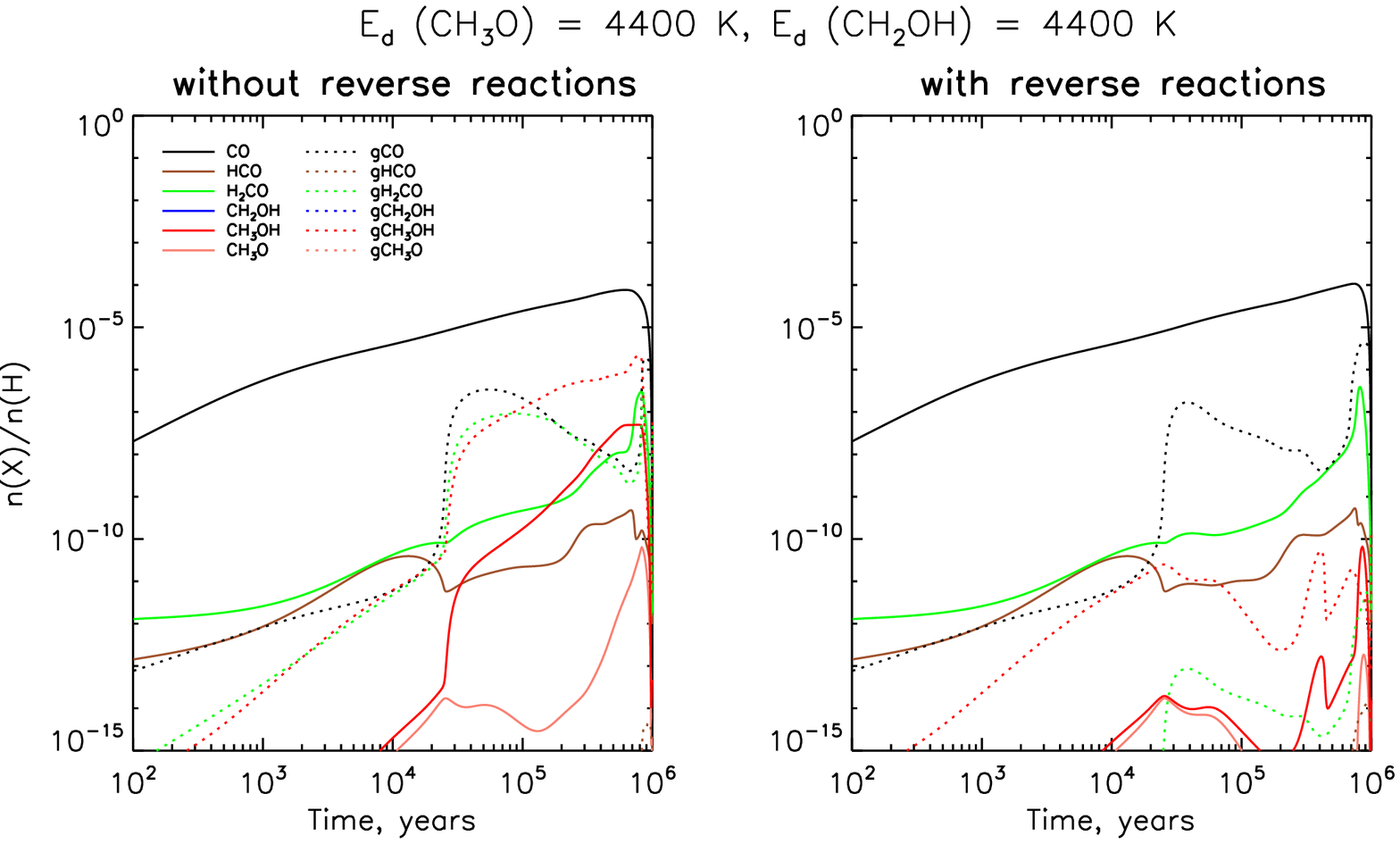}
    \caption{Relative abundances of species versus time for the "cold" stage of the collapse model. These plots correspond to calculations with E$_d$(CH$_2$OH) =  E$_d$(CH$_3$O) = 4\,400~K. Solid line represents molecules in gas phase, dashed line~--- molecules on grain surface. The evolution time range is 10$^2$---10$^6$ years.}
    \label{fig:collapse_cold1}
\end{figure*}

\begin{figure*}
    \includegraphics[width=0.9\textwidth]{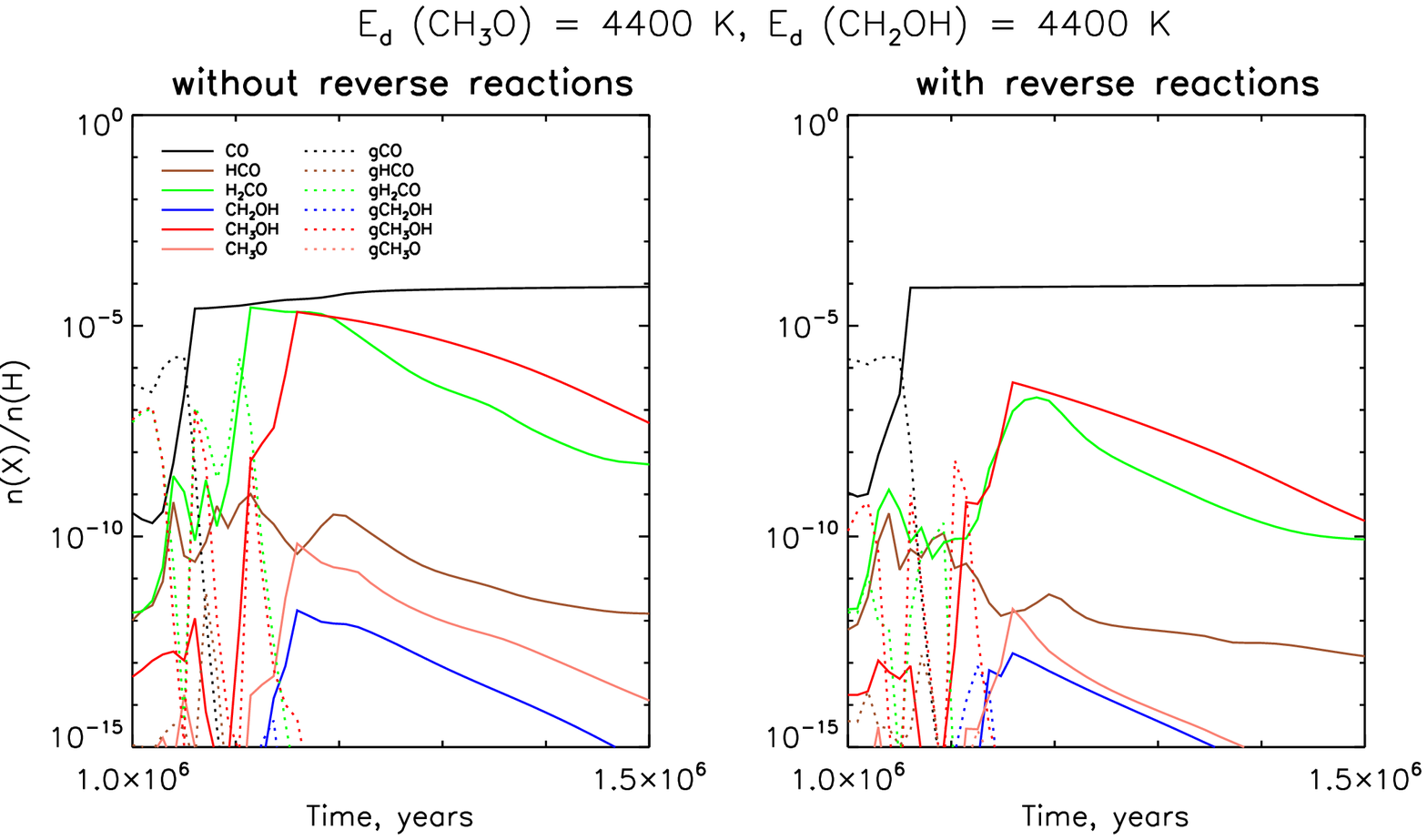}
    \caption{Relative abundances of species versus time for the "warm-up" stage of the collapse model. These plots correspond to calculations with E$_d$(CH$_2$OH) =  E$_d$(CH$_3$O) = 4\,400~K. Solid line represents molecules in gas phase, dashed line~--- molecules on grain surface. The evolution time range is 1---1.5$\times$10$^6$ years.}
    \label{fig:collapse_warm1}
\end{figure*}

Moreover, it is important to study CH$_{3}$O/CH$_2$OH evolution in methanol formation chain. In the recent research of~\cite{Wakelam2017} binding energies for CH$_2$OH and CH$_3$O proposed to be E$_d$(CH$_2$OH) = E$_d$(CH$_3$O) = 4\,400~K. Figures~\ref{fig:coldcloud1}, ~\ref{fig:collapse_cold1} and~\ref{fig:collapse_warm1} show results of calculations for the model of cold dark cloud and for both stages of the collapse from translucent into dark cloud with E$_d$ values from~\cite{Wakelam2017}. Here we will name these cases as Model Ia for model without reverse reactions and new E$_d$ values, and Model IIa for model with reverse reactions and new E$_d$ values. In case of the Model Ia, the most volatile molecules are CH$_{3}$O and CH$_{3}$OH. Such changes CH$_{3}$OH abundances probably can be explained by decreasing of reaction~\eqref{eq: reaction3} rate.

In both stages of the collapse of translucent cloud into dark cloud the only molecules which abundances noticeably changed in both models (in comparison with Model I and II) are CH$_{2}$OH (both gaseous and on grains) and CH$_{3}$O in gas phase. Analysis of its formation pathways shows that changes in binding energies of CH$_{2}$OH and CH$_{3}$O affect only on efficiencies of reactions which were already described above. Therefore, the exact values of binding energies of CH$_{2}$OH and CH$_{3}$O does not affect on the way of methanol formation.

\begin{acknowledgements}
The reported study was funded by RFBR according to the research project 18-32-00645. The author thanks A. B. Ostrovskii and A. I. Vasyunin for valuable discussions during the course of this work. The author thanks the anonymous referee for his helpful review of this paper.
\end{acknowledgements}

\label{lastpage}

\end{document}